\documentclass[twocolumn,twoside, slac_two, floatfix]{revtex4}
\usepackage{graphicx}
\usepackage{fancyhdr}
\begin{document}
\title{The Search for the Sources of the Cosmic Rays One Century after their Discovery}

\author{Francis Halzen}

\affiliation{Department of Physics, University of Wisconsin, Madison, WI 53706,USA}

\begin{abstract}
Despite their discovery potential touching a wide range of science, construction of TeV gamma-ray telescopes, Auger, IceCube and a suite of other particle astrophysics experiments have been largely motivated by the hunt for the sources of cosmic rays. I will assess the status of our search for the still-enigmatic sources of cosmic rays. Although a resolution is decidedly anticipated, the mystery of their origin remains unresolved.
\end{abstract}
\maketitle
\thispagestyle{fancy}

\section{The Sources of the Cosmic Rays: Two Puzzles}
\vspace{.2cm}

\begin{figure}
\centering
\includegraphics[trim =1 340 1 1, clip,width=0.9\columnwidth]{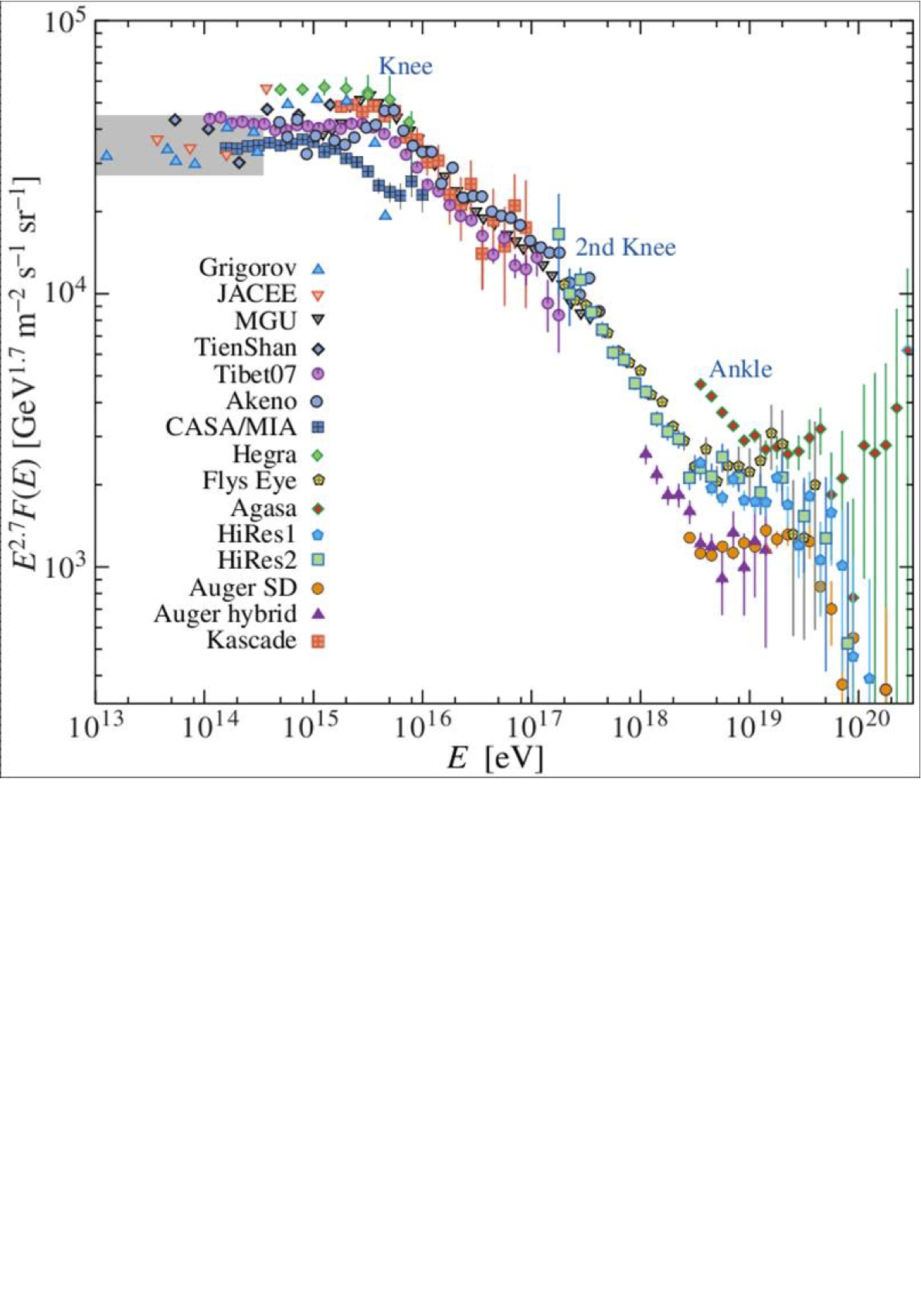}
\caption{At the energies of interest here, the cosmic-ray spectrum
follows a sequence of 3 power laws. The first 2 are separated by the
``knee'', the 2nd and 3rd by the ``ankle''. Cosmic
rays beyond the ankle are a new population of particles produced in
extragalactic sources. Note that the spectrum F(E)(=dN/dE) has been multiplied by a power $E^{2.7}$ in order to visually enhance the structure in the spectrum.}
\label{fig:crspectrum}
\end{figure}

Cosmic accelerators produce particles with energies in excess of $10^8$\,TeV; we still do not know where or how\,\citep{Sommers:2008ji, Hillas:2006ms, Berezinsky:2008qh}. The flux of cosmic rays observed at Earth is shown in Fig.\ref{fig:crspectrum}. The energy spectrum follows a sequence of three power laws. The first two are separated by a feature dubbed the ``knee'' at an energy\footnote{We will use energy units TeV, PeV and EeV, increasing by factors of 1000 from GeV energy.} of approximately 3\,PeV.  There is evidence that cosmic rays up to this energy are Galactic in origin.  Any association with our Galaxy disappears in the vicinity of a second feature in the spectrum referred to as the ``ankle"; see Fig.\ref{fig:crspectrum}. Above the ankle, the gyroradius of a proton in the Galactic magnetic field exceeds the size of the Galaxy, and we are witnessing the onset of an extragalactic component in the spectrum that extends to energies beyond 100\,EeV. Direct support for this assumption now comes from two experiments \,\citep{Abbasi:2007sv, Abraham:2008ru} that have observed the telltale structure in the cosmic-ray spectrum resulting from the absorption of the particle flux by the microwave background, the so-called Greisen-Zatsepin-Kuzmin (GZK) cutoff. The origin of the flux in the intermediate region covering PeV-to-EeV energies remains a mystery, although it is routinely assumed that it results from some high-energy extension of the reach of Galactic accelerators.

\begin{figure}
\centering
\includegraphics[trim =2 425 2 2, clip, width=0.9\columnwidth]{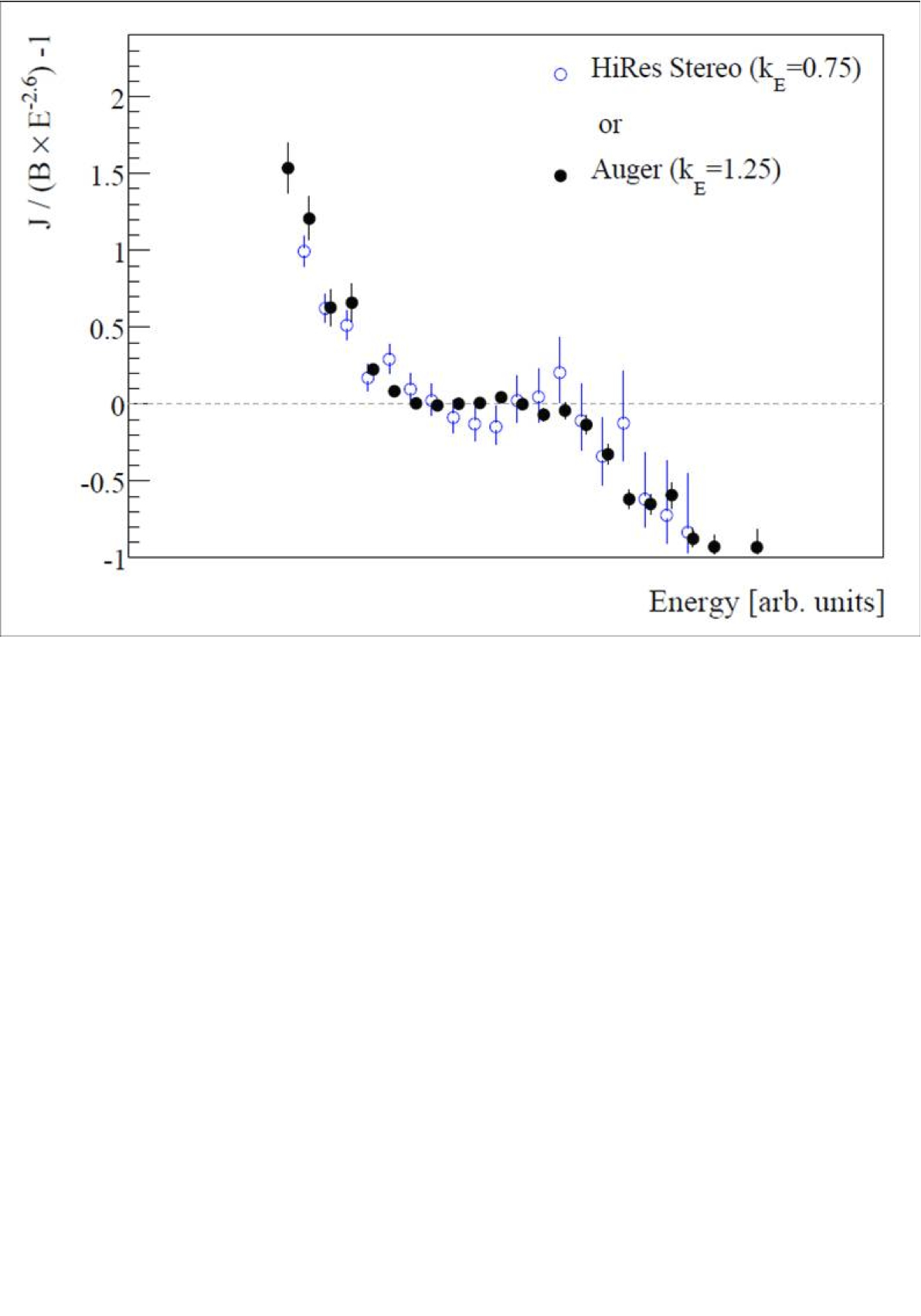}
\caption{Auger and HiRes fluxes agree in the energy range of the GZK feature after a shift which is within the systematic errors of each experiment.}
\label{fig:GZK}
\end{figure}

The agreement between the measured HiRes\,\citep{sokolsky} and Auger\,\citep{privitera} spectra in the GZK energy range is remarkable once one allows for the $\sim\!\! 25\%$ systematic errors on the energy measurement; see Fig.\ref{fig:GZK}. The spectrum has also been confirmed by the very first data from the Telescope Array; it represents a timely addition to the HiRes and Auger experiments that have produced conflicting claims on other properties of the highest energy cosmic rays, most notably their arrival directions and chemical composition\,\citep{fukushima}. We will return to this issue later on but emphasize that the measurements shown in Fig.\ref{fig:GZK} represent an impressive achievement made possible by the renewed investment in cosmic ray detectors in the last decade.

Nevertheless, the origin of the highest energy cosmic rays remains a mystery. The origin of the Galactic cosmic rays is also a mystery-- even though the speculation that they originate in supernova remnants is textbook material, there is no observational evidence for this hypothesis.  

%This is also the case for the Galactic cosmic rays despite the fact that the speculation that they originate in supernova remnants is textbook material for which, in fact, there is no observational evidence.

\section{Best Buy Theory: Follow the Energy!} 

\begin{figure}
\centering
\includegraphics[width=0.9\columnwidth]{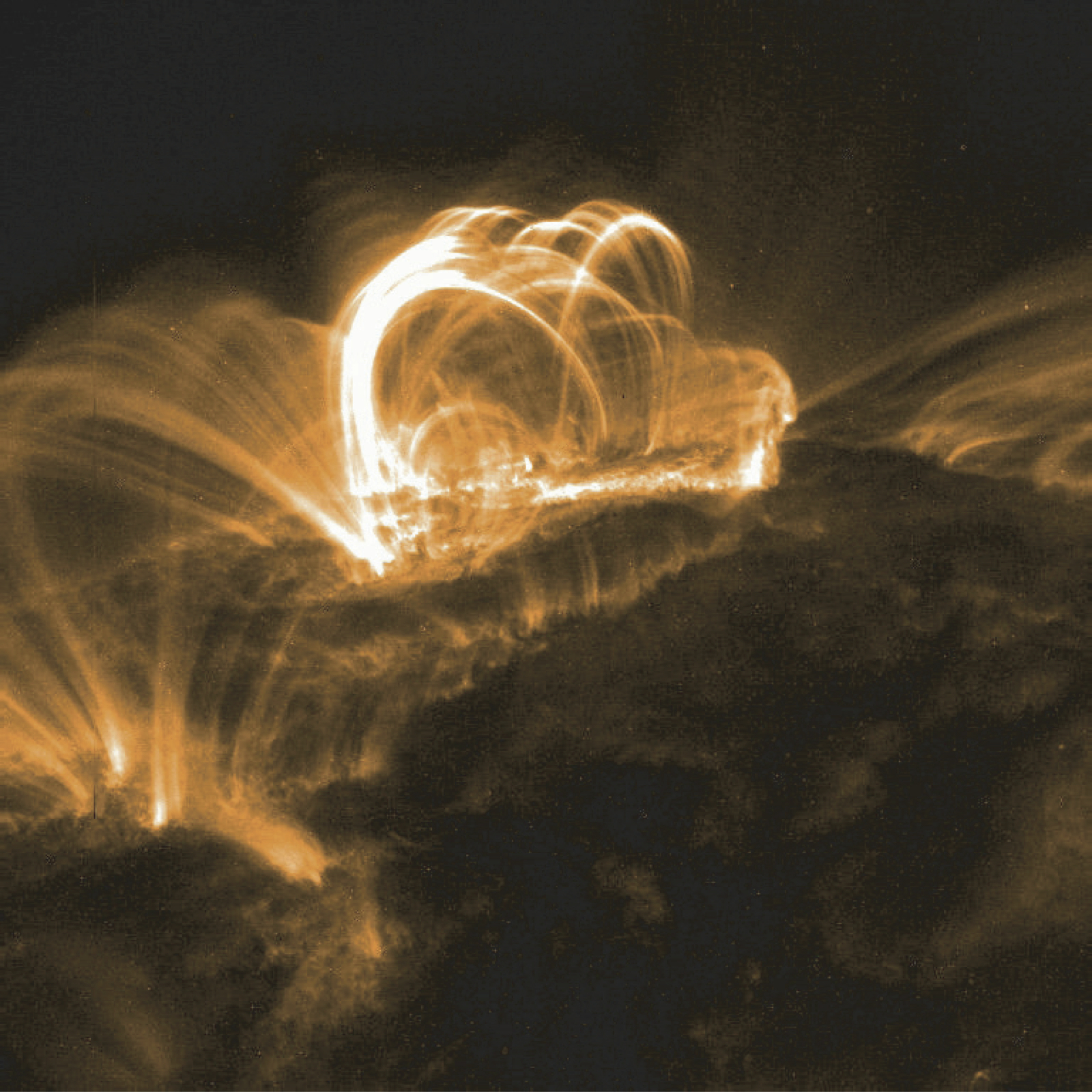}
\caption{Near intense charged particle flows, seen as filaments in this X-ray picture of a solar flare, the opportunity exists for the acceleration of solar particles to GeV energy.}
\label{fig:solar}
\end{figure}

Acceleration of protons (or nuclei) to TeV energy and above requires massive bulk flows of relativistic charged particles. The blueprint of the accelerator can be copied from solar flares where particles are accelerated to GeV energy by shocks and, possibly, reconnection; see Fig.\ref{fig:solar}. Recalling the Hillas formula that states that the gyroradius of the accelerated particle must be contained within the accelerating B-field region, or
\begin{equation}
E \leq Ze\,c\,B\,R,
\end{equation}
reaching TeV energy in solar flares is dimensionally impossible. In a solar flare the extent R of the accelerating region and the magnitude of the magnetic fields B are not large enough to accelerate particles of charge Ze to energies beyond GeV; their velocity is taken to be the speed of light. Particle flows reaching TeV energy and above are likely to originate from exceptional gravitational forces in the vicinity of black holes or neutron stars. Gravity powers large currents of charged particles that are the origin of high magnetic fields. These create the opportunity for particle acceleration by shocks. It is a fact that electrons are accelerated to high energy near black holes; astronomers detect them indirectly by their synchrotron radiation. Some must accelerate protons because we observe them as cosmic rays.

\begin{figure}
\centering
\includegraphics[trim =0 400 2 50, clip,width=0.9\columnwidth]{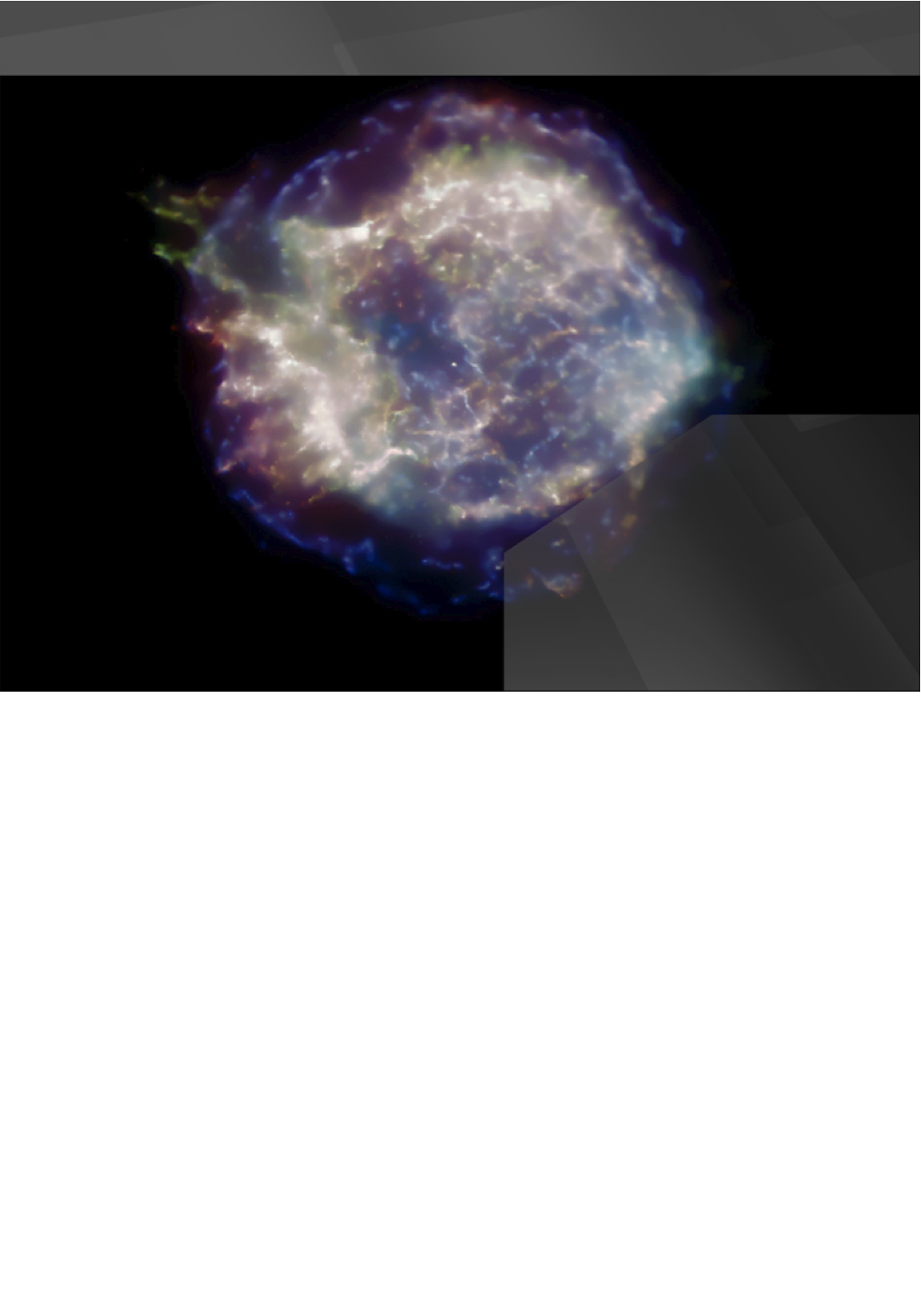}
\caption{This X-ray picture of the supernova remnant CasA reveals strong particle flows near its periphery. We believe they are the site for accelerating Galactic cosmic to energies reaching the ``knee" in the spectrum.}
\label{fig:casA}
\end{figure}

\begin{figure}
\centering
\includegraphics[trim =2 395 2 2, clip,width=0.9\columnwidth]{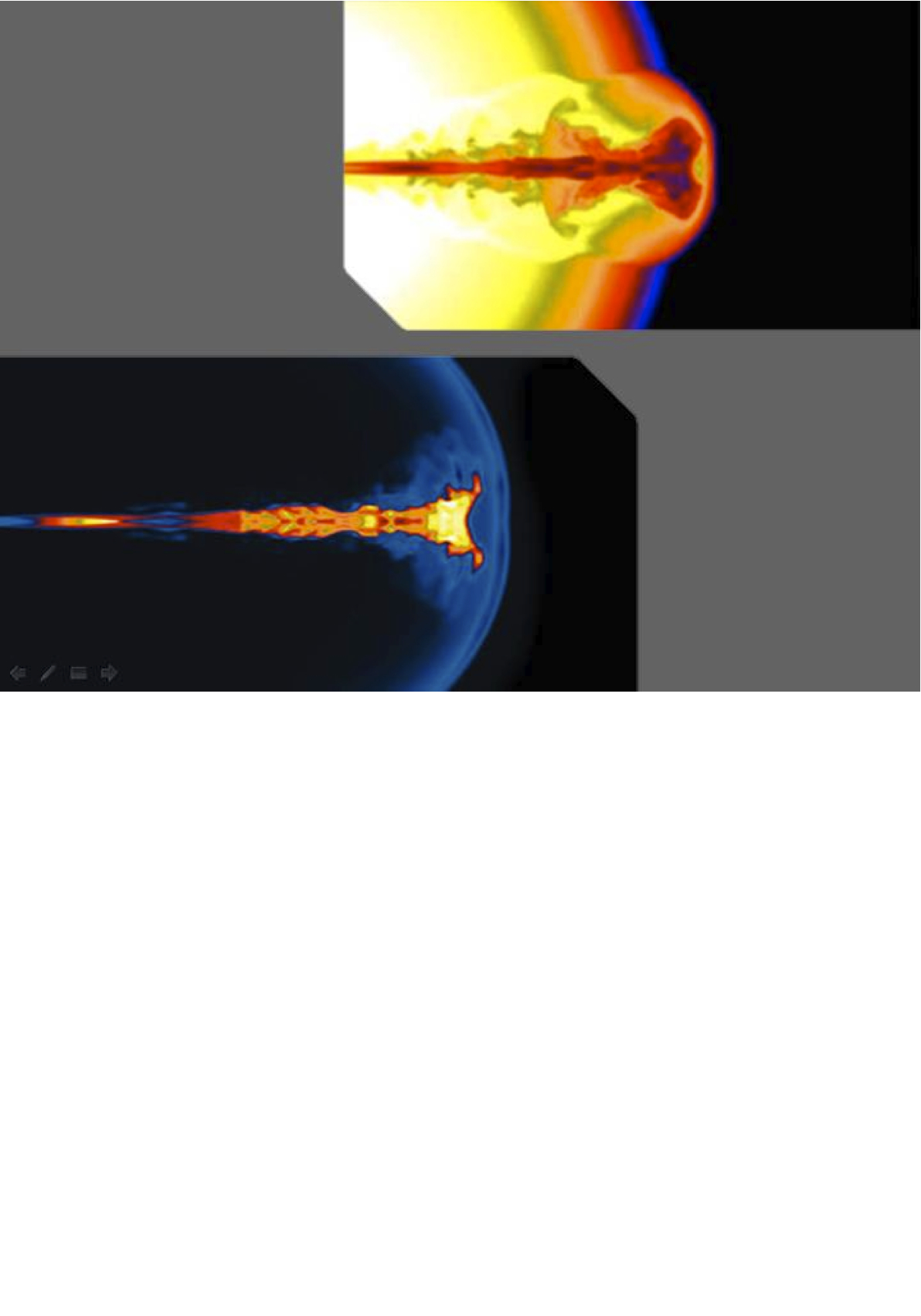}
\caption{Colliding shocks in the simulation of a GRB fireball may accelerate cosmic rays to the highest energies observed. The filaments in the particle flow are directed along the rotation axis of the black hole. Animated view at http://www.nasa.gov/centers/goddard/\\news/topstory/2003/0618rosettaburst.html.}
\label{fig:GRBsim}
\end{figure}

\citet{zwicky} suggested as early as 1934 that supernova remnants could be sources of the Galactic cosmic rays. It is assumed that the accelerators are powered by the conversion of $10^{50}$\,erg of energy into particle acceleration by diffusive shocks associated with young ($\sim\!\! 1000$ year old) supernova remnants expanding into the interstellar medium \,\citep{ADV}. Like a snow plough, the shock sweeps up the $\sim\!\! \,1\,\rm proton/cm^3$ density of hydrogen in the Galactic plane. The accumulation of dense filaments of particles in the outer reaches of the shock, clearly visible as sources of intense X-ray emission, are the sites of high magnetic fields; see Fig.\ref{fig:casA}. It is theorized that particles crossing these structures multiple times can be accelerated to high energies following an approximate power-law spectrum $dN/dE \sim\!\!  E^{-2}$. The mechanism copies solar flares where filaments of high magnetic fields accelerate nuclear particles to tens of GeV. The higher energies reached in supernova remnants are the consequence of particle flows of much larger intensity powered by the gravitational energy released in the stellar collapse.

This idea has been widely accepted despite the fact that to date no source has been conclusively identified, neither of cosmic rays nor of accompanying gamma rays and neutrinos produced when the cosmic rays interact with Galactic hydrogen. Galactic cosmic rays reach energies of at least several PeV, the ``knee" in the spectrum; therefore their interactions should generate gamma rays and neutrinos from the decay of secondary pions reaching hundreds of TeV. Such sources, referred to as PeVatrons, have not been found. Nevertheless Zwicky's suggestions has become the stuff of textbooks and the reason is energetics.

The particles may be few, but each carries a large amount of energy. We derive the average energy density $\rho_{E}$ of cosmic rays in the Galaxy using the relation that the total flux${}={}$velocity${}\times{}$density, or
\begin{equation}
4\pi \int dE \left\{ E{dN\over dE} \right\} =  c\rho_{E}\,,
\end{equation}
i.e. we simply integrate the Galactic part of the flux in Fig.\ref{fig:crspectrum}. The answer is $\rho_{E} \sim\!\!  10^{-12}$\,erg\,cm$^{-3}$, an energy density roughly equal to that in microwave photons (i.e. 410 photons per cm$^3$ of temperature 2.7 K), to that in starlight as well as in the magnetic field. Galactic cosmic rays are not forever; they diffuse within the microgauss fields and remain trapped for an average containment time of  $3\times10^{6}$\,years. The power needed to maintain a steady energy density requires accelerators delivering $10^{41}$\,erg/s. This happens to be 10\% of the power produced by supernovae releasing $10 ^{51}\,$erg every 30 years ($10 ^{51}\,$erg correspond to 1\% of the binding energy of a neutron star after 99\% is lost to neutrinos.) This coincidence is the basis for the idea that shocks produced by supernovae exploding into the interstellar medium are the accelerators of the Galactic cosmic rays.

What about the extragalactic component? When the supernova's predecessor star is sufficiently massive for the collapse to result in a black hole, the remnant will explode into a gamma-ray burst (GRB) fireball where the opportunity exists to accelerate particles by shocks. Like in a  supernova remnant but with a difference, it takes seconds instead of a thousand years; see Fig.\ref{fig:GRBsim}\,\citep{GRBA, Bottcher:1998qi,GRBB}. We can calculate the energy density of the highest energy cosmic rays in the Universe on the back of an envelope. Their flux is at the level of one particle per kilometer squared per year for a typical array with an angular acceptance of one steradian. This can be translated into an energy flux
\begin{eqnarray}
E \left\{ E\frac{dN}{dE} \right\} &=& \frac{10^{19}\,{\rm eV}} {\rm (10^{10}\,cm^2)(3\times 10^7\,s) \, sr} \nonumber \\
&=& 3\times 10^{-8}\rm\, GeV\ cm^{-2} \, s^{-1} \, sr^{-1} \,.
\label{crflux}
\end{eqnarray}
After integrating the flux, as before, we obtain the energy density
\begin{equation}
\rho_{E} = {4\pi\over c} \int_{E_{\rm min}}^{E_{\rm max}} {3\times 10^{-8}\over E} dE \, {\rm {GeV\over cm^3}} \simeq 10^{-19} \, {\rm {TeV\over cm^3}} \,,
\end{equation}
taking the extreme energies of the accelerator(s) to be $E_{\rm max} / E_{\rm min} \simeq 10^3$. The energy content derived ``professionally" by integrating the observed extragalactic spectrum in Fig.\ref{fig:crspectrum}, including the GZK feature, is $\sim\!\! \,3 \times 10^{-19}\rm\,erg\ cm^{-3}$ \,\citep{TKGA,TKGB}. This is within a factor of two of our back-of-the-envelope estimate (1\,TeV = 1.6\,erg).

A GRB fireball converts a fraction of a solar mass into the acceleration of electrons, seen as synchrotron photons. The observed energy in extragalactic cosmic rays can be accommodated with the reasonable assumption that shocks in the expanding GRB fireball convert roughly equal energies into the acceleration of electrons and cosmic rays. It so happens that $\sim\!\! 2 \times 10^{52}$\,erg per cosmological gamma ray burst will yield the observed energy density in cosmic rays after $10^{10}$ years given that their rate is of order 300 per $\textrm{Gpc}^{3}$ per year. Hundreds of bursts per year over Hubble time produce the observed cosmic ray density, just like 3 supernova per century accommodate the steady flux in the Galaxy. Problem solved? Not really, it turns out that the same result can be achieved with active galaxies\,\citep{ginzberg}.

In fact, the energy density of $3 \times 10^{-19}\rm\,erg\ cm^{-3}$ works out to not only \,\citep{TKGA, TKGB}
\begin{itemize}
\item $\sim\!\!  2 \times 10^{52}$\,erg per cosmological gamma- \\ ray burst, but also to
\item $\sim\!\!  3 \times 10^{42}\rm\,erg\ s^{-1}$ per cluster of galaxies,
\item $\sim\!\!  2 \times 10^{44}\rm\,erg\ s^{-1}$ per active galaxy.
\end{itemize}
The coincidence between above numbers and the observed output in electromagnetic energy of these sources explains why GRB and AGN have emerged as the leading candidates for the cosmic ray accelerators.

\section{Multi-wavelength Astronomy: Cosmic Rays, Gamma Rays and Neutrinos}
\vspace{.2cm}

The basic reason that the sources of the cosmic rays have escaped detection is the failure of charged particles to point back at their sources after propagation through the Galactic magnetic field. While this may not be the case for protons with energies close to $10^{20}$\,eV, their flux is low. Alternatives are to search for gamma rays and neutrinos that are produced in association with the cosmic-ray beam. Generically, a cosmic-ray source should also be a beam dump. Cosmic rays accelerated in regions of high magnetic fields near black holes inevitably interact with radiation (and gas) surrounding them, e.g., UV photons in active galaxies or MeV photons in gamma-ray--burst fireballs. In these interactions, neutral and charged pion secondaries are produced by the processes
\begin{eqnarray*}
p + \gamma \rightarrow \Delta^+ \rightarrow \pi^0 + p
\mbox{ \ and \ }
p + \gamma \rightarrow \Delta^+ \rightarrow \pi^+ + n.
\end{eqnarray*}
While secondary protons may remain trapped in the high magnetic fields, neutrons and the decay products of neutral and charged pions escape. The energy escaping the source is therefore distributed among cosmic rays, gamma rays and neutrinos produced by the decay of neutrons, neutral pions and charged pions, respectively. In the case of Galactic supernova shocks, cosmic rays interact with gas, e.g. with dense molecular clouds, as well as radiation, producing equal numbers of pions of all three charges in hadronic collisions $p+p \rightarrow N\,[\,\pi^{0}+\pi^{+} +\pi^{-}]+X$.

Despite the rapid development of instruments with improved sensitivity, it has been impossible to conclusively pinpoint the sources of cosmic rays by identifying accompanying gamma rays of pionic origin. Separating photons radiated or upscattered by electrons from pionic ones is a challenge that has not been met. Kilometer-scale neutrino detectors have the sensitivity to reveal generic cosmic-ray sources with an energy density in neutrinos comparable to their energy density in cosmic rays\,\citep{TKGA, TKGB} and pionic TeV gamma rays\,\citep{AlvarezMuniz:2002tn}. Detecting the accompanying neutrinos would provide incontrovertible evidence for cosmic ray acceleration in the sources but the construction of the huge detectors required to reach the nominal sensitivity has been a challenge.

\section{Sources of Galactic Cosmic Rays?}
\vspace{.2cm}

Galactic cosmic rays reach energies of at least several PeV, the ``knee" in the spectrum. Their interactions with Galactic hydrogen in the vicinity of the accelerator should generate gamma rays from the decay of secondary pions that reach energies of hundreds of TeV. Such sources should be identifiable by a relatively flat energy spectrum that extends to hundreds of TeV without attenuation; they have been dubbed PeVatrons. Straightforward energetics arguments are sufficient to conclude that present air Cherenkov telescopes should have the sensitivity necessary to detect TeV photons from PeVatrons\,\citep{GonzalezGarcia:2009jc}.

These sources may have been revealed by an all-sky survey in $\sim\!\!10$\,TeV gamma rays with the Milagro detector\,\citep{Abdo:2006fq}. A subset of sources located within nearby star-forming regions are identified; some are not readily associated with known supernova remnants or with non-thermal sources observed at other wavelengths!  Subsequently, directional air Cherenkov telescopes were pointed at three of the sources, revealing them as PeVatron candidates with an approximate $E^{-2}$ energy spectrum that extends to tens of TeV without evidence for a cutoff\,\citep{hesshotspot, Albert:2008yk}, in sharp contrast with the spectra of the best studied supernova remnants RX J1713-3946 and RX J0852.0-4622 (Vela Junior).

Some Milagro sources may actually be molecular clouds illuminated by the cosmic-ray beam accelerated in young remnants located within $\sim\!\! 100$\,pc. One expects indeed that multi-PeV cosmic rays are accelerated only over a short time period, when the remnant transitions from free expansion to the beginning of the Sedov phase and the shock velocity is high. The high-energy particles can produce photons and neutrinos over much longer periods when they diffuse through the interstellar medium to interact with nearby molecular clouds; for a detailed discussion, see~\citet{gabici}. An association of molecular clouds and supernova remnants is expected in star-forming regions.

Despite the rapid development of instruments with improved sensitivity, it has been impossible to conclusively pinpoint supernova remnants as the sources of cosmic rays by identifying accompanying gamma rays of pionic origin. Detecting the accompanying neutrinos would provide incontrovertible evidence for cosmic ray acceleration in the sources.  The particle physics relating charged and neutral pion production cross sections dictates the relation between gamma ray and neutrino fluxes; simple counting of final states predicts the arrival at Earth of a $\nu_\mu+\bar\nu_\mu$ pair for every two gamma rays seen by Milagro. This calculation can be performed in a more sophisticated way with approximately the same outcome. For average values of the parameters describing the flux, the completed IceCube detector could confirm sources in the Milagro sky map as sites of cosmic-ray acceleration at the $3\sigma$ level in about one year and at the $5\sigma$ level in three years\,\citep{GonzalezGarcia:2009jc}. These results agree with previous estimates \,\citep{hkm} but both are subject to uncertainties associated with the less than complete knowledge of the sources and their spectra. In the absence of observation of TeV-energy supernova remnant neutrinos by IceCube, the nature of sources that produce cosmic rays near the knee of the spectrum is likely to remain unresolved until the commissioning of next-generation gamma ray detectors such as HAWC and CTA.

\begin{figure}
\centering
\includegraphics[trim =2 535 2 2, clip,width=0.9\columnwidth]{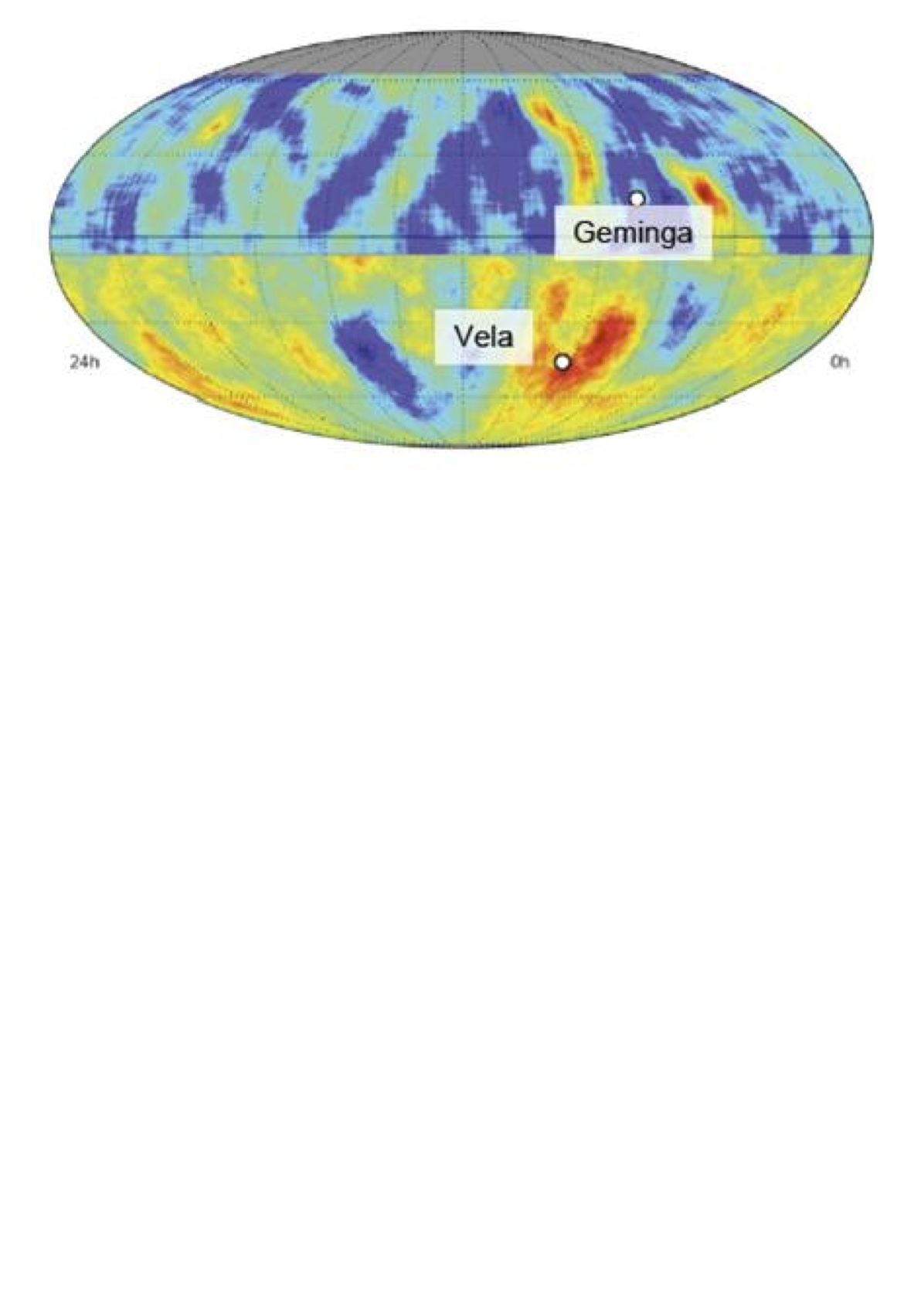}
\caption{The arrival direction of cosmic-ray muons recorded with 40 IceCube strings (Southern Hemisphere). The variations are of order $10^{-3}$ on a uniform distribution. The color scale represents the relative intensity. The dots indicate the directions of Vela, the brightest gamma-ray source in the sky, and Geminga. Also shown is the muon data of Milagro obtained by the same method (Northern Hemisphere).}
\label{fig:muonmap}
\end{figure}

Reminiscent of the discredited Cygnus X-3 observations of the 1980's, muons have rejoined the particle astronomy triad of cosmic rays, gamma rays and neutrinos\,\citep{goodman}. New carpet air shower arrays like the Tibet AS\,$\gamma$ array and ARGO-YBJ as well as Milagro have collected unprecedented statistics on the arrival directions of muons in the $1-10$\,TeV energy range. With IceCube in the Southern hemisphere, they have mapped the arrival direction of the Galactic cosmic rays over the full sky; see Fig.\ref{fig:muonmap}. Using only 4.3 billion downward-going events from half the detector, the IceCube data show an anisotropy in excess of $5\sigma$ in the arrival directions of the cosmic rays\,\citep{Abbasi:2010mf}. The deviation from isotropy is at the 0.1\% level.  The median muon energy is about 20\,TeV and the primary energies of the Galactic cosmic rays that produce them are even higher.   This is a puzzling result, as the arrival directions of charged particles of such energy should be scrambled by Galactic magnetic fields.

Proposed interpretations fall into two categories: the asymmetry in arrival directions of cosmic rays is either associated with unknown structure in the Galactic magnetic field, or with diffusive particle flows from nearby Galactic sources such as Vela, the strongest gamma ray source in the sky; see Fig.\ref{fig:muonmap}. The broad cosmic-ray anisotropy shown in the figure aligns with observations in the Northern Hemisphere. It is intriguing that also here a prominent structure seems to be associated with a nearby photon source, Geminga. Possibly, the sources of the Galactic cosmic rays have thus been revealed in a most surprising way; the challenge will be to confirm this interpretation of the observations with the rapidly accumulating data. Note that it is a challenge, possibly to common sense: the gyroradius of particles with energy exceeding 100 TeV in a microGauss field is less than 0.1 pc compared to the distances to nearby sources such as Vela that exceed 100\,pc.

\section{Sources of the Extragalactic Cosmic Rays?}
\vspace{.2cm}

Whereas one can conceive of a path towards solving the Galactic cosmic ray problem, the situation is murky where the search for extragalactic sources is concerned. The Auger experiment initially observed a correlation of their arrival directions with AGN, although with a 1\% probability of the correlation being due to a fluctuation. It has not been confirmed with increased statistics\,\citep{privitera}. HiRes data show isotropic arrival directions. IceCube, half completed, has performed an analysis searching for neutrinos pointing in the direction of 13 HiRes and 22 Auger events with energy in excess of 57 EeV and  also obtains a 1\% fluctuation probability\,\citep{resconi}. The obvious conclusion is that a lot more statistics is required in order to draw conclusions. The opportunities are several: the Telescope Array that presented first results at this meeting, its extension TALE and Auger North\,\citep{olinto}. The goal is clearly to increase statistics, especially at the highest energies near $10^{20}$\,eV where proton primaries are expected to point back to their sources.

Unlike the case for Galactic cosmic rays, there is no neutrino path to extragalactic sources. Neutrino fluxes from AGN are difficult to estimate; for GRB the situation is qualitatively better provided that they are indeed the sources of the observed cosmic ray flux. Neutrinos of PeV energy should be produced when protons and photons coexist in the GRB fireball\,\citep{Waxman:1997ti}. As previously discussed, the model is credible because the observed cosmic ray flux can be accommodated with the assumption that roughly equal energy is shared by electrons, observed as synchrotron photons, and protons. Calculation of the GRB neutrino flux is straightforward; it is related to the cosmic ray flux by
\begin{eqnarray}
\frac{dN_{\nu}}{dE_{\nu}} &=& \left[1- \left(1-e^{-n_{int}}\right)\right] \frac{1}{3} x_\nu \frac{dN_p}{dE_p}\!\! \left(\!\frac{E_p}{x_\nu}\!\right) f_{GZK} \nonumber\\
& \simeq& n_{int}\, \,x_\nu\,  \frac{dN_p}{dE_p}\!\! \left(\!\frac{E_p}{x_\nu}\!\right),
\label{GRB}
\end{eqnarray}
where $n_{int} \simeq 1$ is the number of interactions of the proton before escaping the fireball; it is determined by the optical depth of the source for $p\gamma$ interactions. Neutrinos reach us from sources distributed over all redshifts, while cosmic rays do so only from local sources inside the so-called GZK radius of less than 100\,Mpc. The evolution of the sources will boost the neutrino flux by the factor $f_{GZK} \simeq 3$ that depends on the redshift distribution of GRB.

Two different and independent searches with the half-completed IceCube detector failed to observe this flux at the 90\% confidence. At this point, rather than ruling out GRB as the sources of the cosmic rays, the issue in front of us is how reliable the calculation is and what range of values of $n_{int}$ and $f_{GZK}$ can be tolerated. The fireball parameters are sufficiently constrained by astronomical observations that $n_{int}$ can be calculated to within a factor of 2\,\citep{Guetta:2000ye}. Knowledge of the enhancement factor $f_{GZK}$ of the neutrino relative to the cosmic ray flux will improve as the number of GRB with known redshift grows; it exceeds unity in any case. Thus, with improved statistics and superior data from the completed instrument, IceCube has the potential in the near future to confirm or rule out GRB as the sources of the highest energy cosmic rays.

\section{Yet More Instrumentation: the LHC}
\vspace{.2cm}

It is clear that the renaissance in the field of cosmic ray physics has been driven by the commissioning of larger and superior instruments, especially air shower arrays and gamma ray telescopes. Our best hope for progress is to pursue this path. As an illustration I will raise the issue that I have avoided so far: chemical composition. My personal opinion is that we do not know it. For Galactic cosmic rays the Kascade experiment\,\citep{arteaga} represented the technology required to obtain the answer and it succeeded, except for the very highest energy particles in the region of the ``knee". The failure is not one of the instrument, it is because there is no adequate theory to simulate the data. The understanding of the particle physics required to infer the properties of the incident cosmic ray from observation at sea level of the shower that it initiated is inadequate. I suspect that this is even more the case for the extragalactic cosmic rays. There are no contradictory measurements; the particle physics to interpret the data is just too uncertain. Therefore, the most critical instrument for the future of cosmic ray physics may be the LHC\,\citep{LHCdiscussion}.

\begin{figure}
\centering
\includegraphics[trim =2 425 2 2, clip,width=0.9\columnwidth]{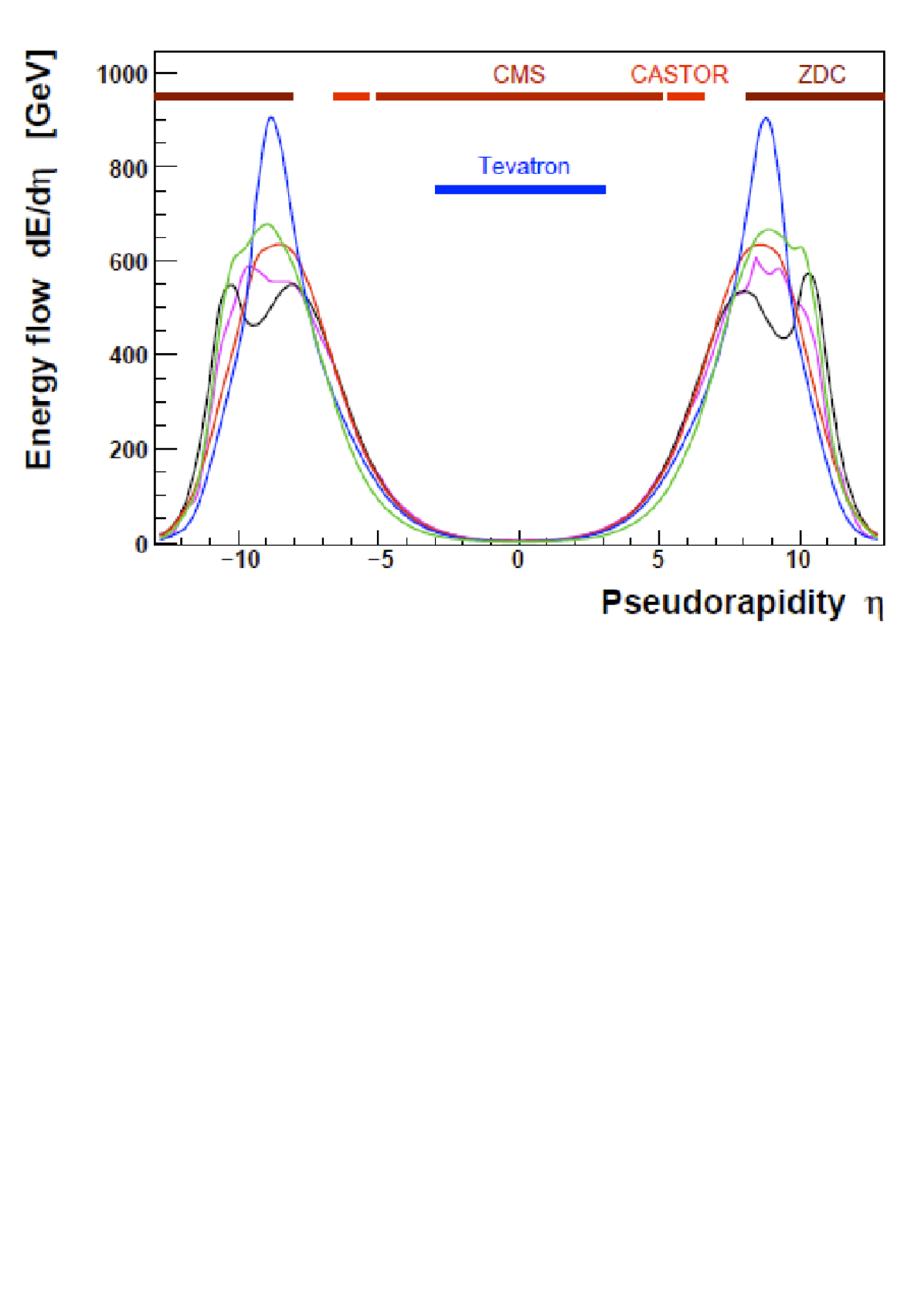}
\caption{Dedicated detectors cover the rapidity range of secondary particles produced in very forward and backward direction at the LHC. The measurements will significantly constrain the particle physics models required to trace observations of cosmic ray showers at sea level back to the properties of the initial particle. Graphics from Ralf Ulrich.}
\label{fig:LHC}
\end{figure}

Acquiring significant measurements  requires the commissioning of dedicated detectors; simple kinematics is sufficient to show that the phase space in rapidity and transverse momentum of secondary particles relevant to the development of air showers is obscured by the beam pipe. This is a technical problem that has been circumvented by ingenious techniques that should deliver relevant data; see Fig.\ref{fig:LHC}. And data is what is required, because extrapolating information from accelerator experiments at the ISR and Tevatron over multiple orders of magnitude is clearly impossible. Let me give one example: the total cross section. For illustration I use SYBILL (or PYTHIA) where the energy dependence of the total cross section is the shadow of the production of high transverse momentum secondaries. Marty Block reminded us at this meeting that this model makes predictions for the LHC and beyond\,\citep{block}. But this is because he assumes that the proton evolves asymptotically into a black disc of gluons, an assumption abandoned by the cosmic ray modelers. The choice of the critical parameter $p_{T}\!\left(\rm{min}\right)$ above which the cross section of secondaries is evaluated drives the rate at which the total cross section grows. In SYBILL this parameter can be arbitrarily chosen at each energy so that the challenging extrapolation between existing data and the energy range of the Auger experiment is no longer constrained by the model. There is no longer any guidance from theory, whether good or bad. The corollary to this argument is that one can freely pick the particle model that gives one's favorite interpretation of the data. The key contribution of the LHC will be to narrow the gap over which information on particle physics will have to be extrapolated.

Given a set of data, can one really change protons into iron? I will only emphasize that we can adjust three critical and free (in the sense just illustrated for the cross section) parameters to tune the models chosen to interpret the data: the cross section, the multiplicity and the inelasticity of the secondaries. While waiting for the relevant LHC data, the best opportunity for progress is possibly to try to understand why HiRes and Auger disagree on the fluctuations of their highest energy events. In the Auger experiment every high energy shower looks like the next one observed, a fact most readily explained by assuming that the energy of an Fe primary is subdivided into constituent nucleons at the top of the atmosphere, thus reducing the fluctuations from event to event. HiRes data exhibit fluctuations consistent with proton primaries. The disagreement could of course be the result of event selection which, again, has to be corrected for using a particle physics model. The disagreement represents an outstanding opportunity for experiments to collaborate on digging into the problem by comparing data and simulations.

\section{Yet More Instrumentation: GZK Neutrinos}
\vspace{.2cm}

The most straightforward way to pinpoint the sources of the highest energy particles is to collect a significant sample of events with energy in excess of $10^{20}$\,eV. We expect that protons will point back to sources within the GZK radius of 75\,Mpc with a resolution of a few degrees.  If the primaries are heavy, we may be out of luck. The most straightforward way to achieve this goal is to build Auger North with a detection increased by a factor of 7 relative to the operating array\,\citep{olinto,privitera}. 

An alternative is to develop new detection techniques. Two promising approaches are being pursued: bistatic radar and microwave detection. An R\&D effort co-located with the Telescope Array is evaluating the potential of using radar to detect the ionization column produced by an ultra-high energy cosmic ray shower in the atmosphere. The key advantage is that the forward Thomson scattering cross section of the radar on free electrons is large. Radar holds the promise to attain greater volume coverage and nearly the accuracy of fluorescence systems with less infrastructure and much longer duty cycle. The estimated mean echo lifetime is on the order of 50 microseconds for a cosmic ray of energy $10^{19}$\,eV. Signals from a continuous wave bi-directional radar system transmitting in the low-VHF with emitter and receiver obscured by a mountain or earth curvature, would map the ionization traced by cosmic rays in the atmosphere. Proof of concept has been achieved by Helio Takai exploiting an outreach project\,\citep{Damazio:2004kh}!

An alternative approach pioneered by Peter Gorham and collaborators exploits the fact that extensive air showers emit signals in the microwave band of the EM spectrum above 1\,EeV. These originate in collisions of the free electrons with the atmospheric neutral molecules in the plasma produced by the passage of the shower. Like the bistatic radar, the technique would allow the longitudinal measurement of an air shower in a manner similar to a nitrogen fluorescence detector, but with a much higher duty cycle and without uncertainty due to variable atmospheric attenuation\,\citep{privitera}.

Finally, there are GZK neutrinos. At energies above $10^{17}$\,eV, a ÒguaranteedÓ source of neutrinos emerges. These GZK neutrinos are produced when cosmic-ray protons with energies above $4\times 10^{19}$\,eV interact with the cosmic microwave background photons. GZK neutrinos point back to their sources. The reason is that a neutrino produced within a 75 Mpc GZK radius from its source located at a typical cosmological distance of the order of Gigaparsecs, will pinpoint its location within the $\sim\!\! 1$\,degree angular resolution of a neutrino telescope. The predicted rate for GZK neutrinos is however of the order of 1 event per km$^3$ per year; IceCube is not big enough to accumulate interesting statistics. One must trade threshold energy for active volume. The coherent Cherenkov emission from electromagnetic showers with energy in excess of $\sim\!\! 10$\,PeV is calculable and, more importantly, verified using the SLAC electron beam striking an ice target. Measurements of the rf power, frequency spectrum, and angular distributions are in agreement with theoretical predictions. In cold ice, the radio-wave attenuation length is of the order of 1 km, an order of magnitude longer than for optical photons; this creates the opportunity to envisage 100\,km$^3$ detectors using a reasonable number of detectors\,\citep{Allison:2009rz}.

Most recently, the Antarctic Impulse Transit Antenna (ANITA) balloon experiment has twice circled Antarctica at altitudes around 35,000 m. Its 32 quad-ridged horn antennas scanned about $10^6\,\rm km^3$ of Antarctic ice. Its threshold to detect GZK neutrinos is not optimal and this requires placing the antennas in or very near the active volume. The first effort in this direction was by the Radio Ice Cherenkov Experiment (RICE) collaboration, who installed dipole antennas in AMANDA holes, and set limits down to $10^{17}$\,eV. An R\&D effort is underway to extend the IceCube array outward by placing antennas in shallow holes. This detector, the Askaryan Radio Array, would ultimately cover $1000~ \rm{km^3}$. The Antarctic Ross Ice Shelf Antenna Neutrino Array (ARIANNA) collaboration pioneered a new approach, placing radio
detectors on the 650-m-thick Ross Ice Shelf in Antarctica. The ice-water interface below the ice shelf is a near-perfect reflector for radio waves. With this reflection, radio waves from downward-going neutrinos will reach the surface, increasing ARIANNAÕs sensitivity. A neutrino experiment large enough to observe several GZK neutrinos per year would complement cosmic-ray experiments such as Auger North; for an extensive review see \citet{Chen:2009ra}.

\section{Conclusions}
\vspace{.2cm}

Obviously there are no conclusions regarding the sources of cosmic rays. There is optimism that the instrumentation is in place to pinpoint them and, if not, we are not short on ideas for better detectors.

\bigskip

\begin{acknowledgments} 
I thank my collaborators in IceCube as well as Concha Gonzalez-Garcia, Aongus \'{O} Murchadha, and Ralf Ulrich for contributing to this talk. This research was supported in part by the U.S. National Science Foundation under Grants No.~OPP-0236449 and  PHY-0354776; by the U.S.~Department of Energy under Grant No.~DE-FG02-95ER40896; by the University of Wisconsin Research Committee with funds granted by the Wisconsin Alumni Research Foundation; and by the Alexander von Humboldt Foundation in Germany.
\end{acknowledgments} 
%\newpage
\bigskip

\bibliographystyle{apsrev}
\bibliography{ISVHECR_final}{}

\end{document}